\documentstyle[12pt]{article}

\author{O.Z.Alekperov\thanks{e-mail: semic@lan.ab.az}\\
Institute of Physics of Academy of Sciences of Azerbaijan Republic \\
370143, Baku, H.Javid avenue, 33}
\title{CAPTURE\ OF\ CARRIERS\ TO\ SCREENED\ CHARGED\ CENTRES\ AND\ LOW\ TEMPERATURE\ SHALLOW IMPURITY\ ELECTRIC\ FIELD\ BREAK DOWN IN SEMICONDUCTORS
}
\date{ 
}
\input tcilatex

\QQQ{Language}{
American English
}

\begin{document}

\maketitle
\begin{abstract}
Free carrier capture by a screened Coulomb potential in semiconductors are
considered. It is established that with decreasing screening radius the
capture cross section decreases drastically, and it goes to zero when $%
r_s=a_B^{*}$. On the basis of this result a new mechanism of shallow
impurity electric field break down in semiconductors is suggested.
\end{abstract}

\section{INTRODUCTION}

For correct consideration of the kinetic, photoelectrical and optical
phenomena in semiconductors and semiconductor structures it is necessary to
take into account the carrier capture by attractive centres. One of these
centres in semiconductors are negatively or positively charged shallow
acceptors or donors, the potential of which is considered usually as a
Coulomb interaction. The capture of carriers by a Coulomb centre in
semiconductors was first considered by Lax [1] and was corrected in [2]. In
[2] the capture theory was developed for small and large concentrations of
impurities. In the first case the capture occurs at isolated centres . In
the second case, which is characterised by an overlap of the effective
capture orbits $(r_T=e^2/\chi kT)$ of neighbouring centres, it was supposed
that the capture takes place in the wells of the potential fluctuations of
impurities. This gives an essentially week dependence of the capture cross
section (CCS) on centres concentration $(\sigma \symbol{126} N_d^{1/6})$
compared with that for isolated centres $(\sigma \symbol{126} N_d)$.
However, he potential of the charged impurity in real semiconductors may be
considered as purely Coulombic in the week doping case only ($N_d^{1/3}\cdot
a_B^{*}\ll 1,$where $N_d$ is the shallow impurities concentration, $a_B^{*}$
is an effective Bohr radius).With increasing of impurity concentration the
potential of charged centre changes from a Coulomb to a Yukawa type
potential as a result of screening by free electrons and charged impurities.

In this work we will consider the capture process in the case of a high free
carrier concentration $n$, when Debye screening of a Coulomb centre takes
place. Such a situation can be realized in semiconductors under the
following circumstances:

-in the case of high impurity concentration and at relatively high
temperatures when $kT$ is comparable with the shallow impurity ionization
energy $\epsilon _i$, so that most of shallow impurities are ionized $(n%
\symbol{126} N_d)$;

-in the case of small as well as high concentrations of impurities and low
temperatures $(kT\ll \epsilon _i)$, if a sufficiently strong electric field
is applied to the semiconductor. As it is known [2-3] the CCS would decrease
under the electric field, and as a result free electron concentration would
increase [4]. As it will be shown in the case of strong free electron
screening the CCS goes to zero.

\section{CAPTURE CROSS SECTION TO SCREENED COULOMB CENTRE}

We consider the capture of free carriers by a potential of the form

\begin{equation}
\label{1}U=-(e/\chi r)\exp (-r/r_s) 
\end{equation}

In (1) $r_s$ is the Debye screening radius, and it must be chosen as $%
r_s=\chi \cdot {\bf E}_F/(6\pi ne^2)$ in the degenerate case and as $r_s= 
\sqrt{\chi kT/(4\pi ne^2)}$ in the nondegenerate case, where ${\bf E}%
_F=h^2k_F^2/2m^{*}$, $k_F=(4\pi n^2)^{1/3},$ $\chi $ is dielectric constant
and $n$ is the free carrier concentration.Note that in the conduction band
bottom of gap semiconductors the carriers distribution can be taken as
Boltzman one also in low temperature and high concentration case.

Similar to Coulomb potential case, the effective capture radius of centre is
determined from the equation 
\begin{equation}
\label{2}{\bf E}=(e^2/\chi r)\cdot \exp (-r/r_s) 
\end{equation}

where ${\bf E}$ is the total energy of the carriers. In contrast to Coulomb
potential case equation (2) is transcendental, and can not be solved
analytically.

To calculate the CCS we use the following expression [2]:

\begin{equation}
\label{3}\sigma =(\pi h)^2/(2kTm^{\star })\left[ \stackrel{0}{\stackunder{%
-\infty }{\int }}\exp ({\bf E}/kT)B^{-1}({\bf E})d{\bf E}\right] ^{-1} 
\end{equation}

where 
\begin{equation}
\label{4}B({\bf E})=\int \epsilon \tau ^{-1}(\epsilon )\rho (\epsilon
)\delta \left[ {\bf E}-\epsilon -U\left( r\right) \right] d\epsilon d^3r 
\end{equation}
\begin{equation}
\label{5}\rho \left( \epsilon \right) =8\sqrt{2}\pi (2\pi h)^{-3}m\star
^{3/2}\epsilon ^{1/2},\tau \left( \epsilon \right) =l_0(m\star /(2\epsilon
))^{1/2},l_0=(\pi h^4\rho _0)/(2m\star ^3{\bf E_c^2}) 
\end{equation}
${\bf E_c}$ is the deformation potential constant,$\rho _0$ is the crystal
density, $m^{\star }$ is the carrier effective mass.At low temperatures
electrons are distributed between the impurity ground state $1s$ up to
conduction band bottom. In such a situation carriers can not be captured by
emission of an optical phonons because theirs energy is greeter than
distances between shallow impurity states (at least for most of
semiconductors). For this reason formula (3) describes capture owing to
diffusion lowering of carriers as a result of their wandering between
excited states of impurity by absorption or emission of acoustic phonons
only.

Substituting (1) and (5) into (4) and after integrating using $\delta $%
-function properties, it is easy to obtain for $B({\bf E})$ an expression. 
\begin{equation}
\label{6}B({\bf E})=8m^{\star }/(\pi l_0h^3)\left[ \frac 13{\bf E}^2r_i^3+2%
{\bf E}^2r_ir_s^2(1+\frac{r_i}{r_s}-e^{\frac{r_i}{r_s}})+\frac 12{\bf E}%
^2r_i^2r_s\left( e^{\frac{r_i}{r_s}}-1\right) e^{\frac{r_i}{r_s}}\right] 
\end{equation}

The expression for $B({\bf E})$ can be written in the form: 
\begin{equation}
\label{6}B({\bf E})=8m\star /(\pi l_0h^3)\cdot r_s^3{\bf E}^2/6\cdot J(x) 
\end{equation}
\begin{equation}
\label{7}J(x)=2x^2+12x(1+x-\exp (-x))+3x^2(\exp (x)-1)\exp (x) 
\end{equation}
where $x=r_i/r_s$, $r_i$ is the root of equation (2) for a given screening
length $r_s$. Note that in obtaining (6) and (7) for each $r_s$ we first
find $r_i$ numerically from (2), and then substitute this value as an upper
limit of the integral (4).

Substituting (6) and (7) into (3) we obtain an expression for CCS

\begin{equation}
\label{8}\frac{\sigma _0}\sigma =2/(kT)^2\cdot (e^2/\chi r_s)^3\stackrel{%
\infty }{\stackunder{0}{\int }}\exp (-{\bf E}/kT)/({\bf E}^2J(x))\cdot d{\bf %
E} 
\end{equation}

where%
$$
\sigma _0=(4\pi /3l_0)\cdot (e^2/\chi kT) 
$$
is the CCS in the Coulomb potential case.

The results of numerical calculation of $\sigma _o/\sigma $ dependence on $%
r_s/a_B^{*}$ at $T=4.2K$ for $GaAs$ (curve 1) and $Ge$ (curve 2) with
parameters $m^{\star }=0.067m$, $\chi =12.5$ and $m^{\star }=0.082m_0$, $%
\chi =16$, respectively, are shown in Fig.1.

It is easy to show that when $r\rightarrow \infty $ for CCS from equation
(8) the Coulomb potential case is obtained. Note, that the screened
potential (1) in contrast to the Coulomb one has finite number of bound
states, and when $r\leq a_B^{\star }$ has no bound states at all -they pass
into the continuous bands [5,6]. It is obvious that in the absence of bound
states the CCS must be equal to zero for such a centre. But as it is seen
from Fig. 1 when $r_{s=}a_B^{\star }$ the CCS in comparison with Coulomb
potential case decreases no more than 20 and 25 times for Ge and GaAs
correspondingly. This means that the diffusive method used for the CCS
calculation in [2] and in this work becomes inapplicable at small screening
lengths, when the number of discrete states is small. In this case the
capture process can not be considered as a diffusive lowering of carriers
through energetic states of impurity. Note that owing to this the values of $%
\sigma _o/\sigma $ would be higher than those presented by curves 2 and 3
not only for $r_{s=}a_B^{\star }$.

Thus we obtain the simple result -the more the screening the less is the
capture coefficient, and when $r_{s=}a_B^{\star }$ it is equal to zero. It
is obvious that the analogous result must be obtained for the coefficient of
thermal ionization from impurity states because of the lowering of the
ionization energy $\epsilon _i$ from them when screening is strong
(ionization probability $w_i\sim \exp (-\epsilon _i/kT)$). Now we will
consider some consequences of the obtained result.

\section{LOW TEMPERATURE SHALLOW IMPURITY ELECTRIC FIELD BREAK-DOWN MECHANISM
}

We will discuss the low temperature shallow impurity electric field
breakdown ( LTSIEFB) phenomenon in semiconductors. From the first
observations of LTSIEFB [7] up to now [8] it is believed that this
phenomenon is only due to impact ionization of neutral impurities by free
electrons as a result of their heating under an external electric field. Our
result allows to put forward an alternative mechanism for LTSIEFB which
explains all peculiarities of current voltage characteristic (CVC) of
semiconductors including avalanche-like increase of current and '' S ''-like
form of CVC at breakdown electric field. According to this mechanism with
increasing of electric field the concentration of free carriers $n$ will
increase because of well knowing decrease of capture coefficient $\alpha $
and increase of ionization coefficient $\beta $.The value of $n$ in electric
field would be established by balance condition between capture and thermic
ionization - $n\alpha N_D^{+}=\beta N_D^o$ ($N_D^o$- neutral and $%
N_D^{+}=N_A+n$-charged donors concentrations)

\begin{equation}
\label{9}n(E)=N_D^o(E)/N_D^{+}(E)\cdot \beta (E)/\alpha (E) 
\end{equation}

At some electric field, which is very close to the breakdown one, the value
of $n$ would be so much that the screening of charged impurities will take
place. From this moment an avalanche increase of free carrier concentration
will begin, owing to the CCS decrease because of screening and, as a result
of this, a further increase of $n(E)$, and so on. Thus $n(E)$ and as well as

\begin{equation}
\label{10}j(E)=en(E)\mu (E)E 
\end{equation}

-dependencies will show an avalanche-like increase with electric field. Note
that LTSIEFB takes place at low temperatures when the dominant scattering
mechanism of carriers are charged impurities. This means that owing to the
screening of charged impurity potentials, the mobility of carriers $\mu (E)$
at the breakdown electric field will increase, and as a result of this the
CVC would have an ''S''-like character. Screening induced $\mu (E)$ increase
causes an additional (besides of $n(E)$) current increase in the
avalanche-like region of the CVC. Note that it was already established from
cyclotron resonance line shape investigations of $n-GaAs$ that free carrier
screening of charged impurities is strong at the breakdown electric fields
[9-10].For LTSIEFB there is no need for the condition $r_s=a_B^{*}$, when
total screening of impurity state occurs. First of all such a condition
means all neutral shallow impurities to be ionized in semiconductor. But as
it was shown from Hall measurements [11] at breakdown electric field in $%
n-Ge $ only 5\% and, from plasma shift of cyclotron resonance line in $%
n-GaAs $ [12] at electric fields 3-times greater than the breakdown one,
only about 40\% of neutral impurity were ionized. On the other hand the
condition $r_s=a_B^{*}$ corresponds also to a Mott transition which occurs
at sufficiently high impurity concentrations -$N_D^{1/3}a_B^{*}\approx 0,25$
and in this case all impurity electrons are in the conduction band [13].
Hence LTSIEFB must disappear at very high impurities concentrations. Note
that according to the screening mechanism of LTSIEFB it must disappear in
the low impurity concentration case too, which can be determined from the
condition $r_s=r_T=e^2/\chi kT$. Consequently, according to the supposed
mechanism, LTSIEFB takes place only at neutral impurity concentrations $%
(\chi kT/e^2)^3\cdot (1/4\pi )<N_D^o<(0.25/a_B^{*})^3$. For $n-GaAs$ this
condition requires $5\cdot 10^{11}cm^{-3}<N_D^o<2\cdot 10^{16}cm^{-3}$. In
the next work I will present an experimental evidence which contradicts the
impact ionization model and confirms above mechanism of LTSIEFB in $n-GaAs$.
The fact that the CCS goes to zero when $r_s\leq a_B^{*}$ may be considered
as one of reasons for a Mott transition in semiconductors.

I wish to thank Dr. T.G.Ismailov for calculating and plotting of Fig.1

\newpage\ 

CAPTIONS:

depen.gif

Fig. 1. Dependence of s0/s on screening radius rS/aB*: 1 - for GaAs; 2 - for
Ge; 3 - Coulomb potential case.

\end{document}